\documentclass[fleqn,twoside]{article}
\usepackage{espcrc2}

\usepackage{graphicx}
\usepackage[figuresright]{rotating}

\newcommand{\AmS}{{\protect\the\textfont2
  A\kern-.1667em\lower.5ex\hbox{M}\kern-.125emS}}

\title{Anisotropy studies around the galactic center}

\author{Antoine Letessier-Selvon\address[CBPF]{Centro Brasileiro de Pesquisas Fisicas, Rua Xavier Sigaud 150, Rio de Janeiro, 22290-180, Brazil}
        \thanks{On leave from Laboratoire de Physique NuclŽaire et des hautes énergies, T33 RdC, 4 place Jussieu, 75252 Paris Cedex 05, France}
	for the Auger Collaboration}
       
\begin{document}

\begin{abstract}
We present the first results for anisotropy searches around the galactic center at EeV energies 
using data from the Pierre Auger Observatory. Our analysis, based on a substantially larger data set,
do not support previous claim of anisotropy found in this region by the AGASA and Sugar experiment. 
Furthermore we place un upper limit on a possible point like source located at the galactic center 
which exclude several scenarios predicting neutron sources in this location.

\vspace{1pc}
\end{abstract}

\maketitle

\section{Introduction}
The Galactic Centre (GC)is  an attractive target for the studies of  
cosmic ray (CR) anisotropy at EeV energies.
The GC contains objects that might be candidates for powerful CR accelerators. 
and the  recent high significance observation by H.E.S.S. 
of a TeV $\gamma$ ray source near the 
location of  Sagittarius $A^*$ \cite{hess}, 
together with the discovery of  extended emission from
giant molecular clouds in the central 200~pc of the Milky Way \cite{hess2}, 
motivates the search for excesses in this direction. 
The GC passes only $6^\circ$ from the
zenith at the site  of the Pierre Auger Observatory in the southern hemisphere 
and makes it particularly suitable
for anisotropy studies in this region.
The number of CRs of EeV energies accumulated so
far at  the Pierre Auger Observatory 
from this part of the sky greatly
 exceeds that from previous observations.
We  adopt hereafter for the GC coordinates the 
Sagittarius~$A^*$ J2000.0 coordinates,  $(\alpha,\delta)=(266.3^\circ,-29.0^\circ)$, 

There have also been reports by the 
AGASA experiment \cite{ha98,ha99} 
indicating a $4.5\sigma$ excess of cosmic rays 
 in the energy range $10^{18}$--$10^{18.4}$~eV and by the Sugar
experiment\cite{be01}.
Note that the GC itself lies outside the AGASA field of view ($\delta>-24.2^\circ$).

We also searched for a point-like excess from the GC, as
 EeV  neutrons emitted by one of the possible energetic sources
in the centre of the Galaxy may reach the Earth before decaying, 

After a brief description of the Auger detector we discuss the angular 
resolution of our surface array and the sky coverage estimation. We follow 
with a presentation of the data set used in this analysis and with the results concerning
the AGASA and SUGAR claims as well as an upper limit on a neutron source located at the centre of
our galaxy.

\section{The Auger detector. The hybrid design}
A fundamental characteristics of the Pierre Auger Observatory\cite{ea04} is its capability of \textit{hybrid} reconstruction of cosmic ray
showers~\cite{Mostafa:2005kd}.
Two independent detectors, the Surface Detector (SD), which samples the shower particles at ground, and the Fluorescence Detector (FD),
which collects the fluorescence light emitted by the shower particles along their path in the atmosphere,
are able to measure the energy and direction of the same cosmic ray shower.
The enhanced capabilities of the Auger hybrid detector are examined in this paper.

The Pierre Auger Observatory
was designed to observe, in coincidence, the shower particles at ground
and the associated fluorescence light generated in the atmosphere.
This is
achieved with a large array of water Cherenkov detectors coupled with
air-fluorescence detectors that overlook the surface array.
It is not simply a dual experiment.
Apart from important cross-checks and measurement redundancy,
the two techniques see air showers in complementary ways.
A single air shower is detected 3-dimensionally.
The ground array measures the 2-dimensional lateral structure of the shower at ground level,
with some ability to separate the electromagnetic and muon components.
The fluorescence detector records the longitudinal pro{f}ile of the shower during
its development through the atmosphere.

A \textit{hybrid} event is an air shower that is simultaneously detected
by the fluorescence detector and the ground array.
The Observatory was originally designed and is currently being built with a \textit{cross--triggering}
capability.
Data are recovered from both detectors whenever either system is triggered.
If an air shower independently triggers both detectors the event is tagged
accordingly.
There are also cases where the fluorescence detector,
having a lower energy threshold,
promotes a sub--threshold array trigger.
Surface stations are then matched by timing and location.
This is an important capability because these sub--threshold hybrid events
would not have triggered the array otherwise.
The geometrical reconstruction of the air shower's axis is accomplished by minimizing a
$\chi^{2}$ function involving data from all triggered elements in the eye and on the ground.
The reconstruction accuracy is far better than the ground array counters and the single eye
could achieve independently~\cite{angres}.

The combination of the air fluorescence measurement and the particle detection on the ground provides
an absolute energy calibration.

\section{Surface Array Angular Resolution}
\label{SD-angres}
The angular resolution for the SD is determined, on an event by
event basis, from the zenith ($\theta$) and azimuth ($\phi$)
uncertainties obtained from the geometrical reconstruction, using
the relation:
\begin{equation}
F(\eta) = 1/2~(V[\theta] + \sin^2({\theta})~V[\phi])
\label{eq:ra}
\end{equation}
\noindent
where $\eta$ is the space-angle, and $V[\theta]$ and $V[\phi]$ are the
variance of $\theta$ and $\phi$ respectively. If $\theta$ and
$\phi/\sin(\theta)$ have Gaussian distribution with variance
$\sigma^2$, then $F(\eta) = \sigma^2$ and $\eta$ has a distribution
proportional to $e^{-\eta^2/2\sigma^2}~d(\cos(\eta))d\phi$. Then, if
we define the angular resolution ($AR$) as the angular radius that
would contain 68\% of showers coming from a point source,
$AR~=~1.5~\sqrt{F(\eta)}$.

The angular resolution depends strongly on the timing resolution of
the water Cherenkov detectors (WCDs) and weakly on the shower front
model and the core position uncertainty. The WCDs
timing uncertainty is directly modeled from the data
(section~\ref{model}). This model is
based on the physics of the shower and the measurement process. It can be
adjusted using two pairs of adjacent stations located in the surface
array (section~\ref{doublets}). The model is validated by studying the
$\chi^2$ probability distribution for the geometrical reconstruction
(section~\ref{prob}). The angular resolution is estimated for the
SD-only reconstruction and by comparison with the hybrid data
(section~\ref{SDonly}). 

\subsection{The Time Variance Model}\label{model}

The angular accuracy of the SD events is driven by the accuracy with
which one can measure the arrival time ($T_s$) of the shower front
in each station. The particle arrival time in the shower front can be
described as a Poisson process over some interval time $\mathcal
T$. The first particle arrival time is used as the estimator for the
shower front arrival. It is given by\footnote{ In fact, an
unbiased estimator should be ${T_0} = T_1 - \hat{E[t_1]}$,
where $\hat{E[t_1]}$ is the expectation value.} $T_1 = T_s + t_1$,
where $T_s$ is the shower front time and $t_1$ is taken to follow
an  exponential distribution function with decay parameter ${\mathcal \tau}$.

Since we estimate the parameter $\mathcal T$ from the data itself,
the variance of $T_1$ (given by the variance of $t_1$) is:
\begin{equation}
  V[T_1] = \left( \frac {\mathcal T}{n}\right )^2~
  \frac {n-1}{n+1}
  \label{eq:var0}
\end{equation}

The variance of the arrival of the first particle in the SD stations, 
taking into consideration the GPS uncertainty and the resolution of 
the flash analog-to-digital converters (FADC), can then be written 
as:
\begin{equation}
  V[T_1] = a^2~\left(\frac {2~T_{50}}{n}\right)^2~\frac {n-1}{n+1} + b^2
  \label{eq:var}
\end{equation}
\noindent
where $T_{50}$ is the time interval that contains the first 50\% of
the total signal as measured by the photomultiplier FADC traces. The 
two free parameters $a$ and $b$ can be determined with the adjacent
station data. We expect that the
parameter $a$ should be close to 1, while $b$ should be given by the 
GPS clock accuracy (about 10 ns) and the FADC trace resolution
25/$\sqrt{12}$ ns, that is $b\simeq 12$ ns.

To calculate the number of particles ($n$) we assume that all
particles hit the detector with the same direction than the shower
axis, and that the muons are mostly the ones that contribute to the
time measurements. Then, we obtain $n$ as the ratio between the
total signal ($S$) in the WCD and the average track length,
$TL(\theta)$, of the particles.

The average track length can be computed as the ratio of the detector
volume ($V$) and the area ($A$) subtended by the arriving particles,
and is:
\begin{equation}
  TL(\theta) = \frac{V}{A} = \frac{\pi r^2 h}
  {\pi r^2 \cos(\theta) + 2 r h \sin(\theta)}
\label{eq:fittl}
\end{equation}
\noindent
where $\theta$ is the zenith angle, $r$ = 1.8~m is the detector
radius, and $h$ = 1.2~m is the detector height.

\subsection{Testing the model with doublets}\label{doublets}

Two pairs of adjacent surface detector stations separated by 11~m
(``doublets'') have been installed in the field of the Auger
Observatory. These pairs enable
comparison of timing and signal accuracy measurements.
We used the data of the doublets to verify the time variance model
and also to adjust the constants $a$ and $b$ from it. For each
event we computed the time difference as $\Delta T = dT^{(1)} - dT^{(2)}$
where $dT^{(1)}$ ($dT^{(2)}$) is the time difference from the first (second)
detector of the doublet to the fitted shower front. Doing that,
$\Delta T$ does not depend on the shower front shape, since the twin
detectors are very close to each other. We used 1693 events (from
April/2004 to June/2006) to fit for the two parameters $a$ and $b$,
and we obtained:
\[
\begin{array}{ccrcl}
  a^2&=&0.98 &\pm& 0.05
  \\
  b^2&=&150~\mbox{\rm ns}^2 &\pm& 18~\mbox{\rm ns}^2
\end{array}
\]
which is in good agreement with our expectations
($a^2$ = 1, $b^2$ = 144 ns$^2$).

\begin{figure}[!]
    \includegraphics*[width=0.45\textwidth]{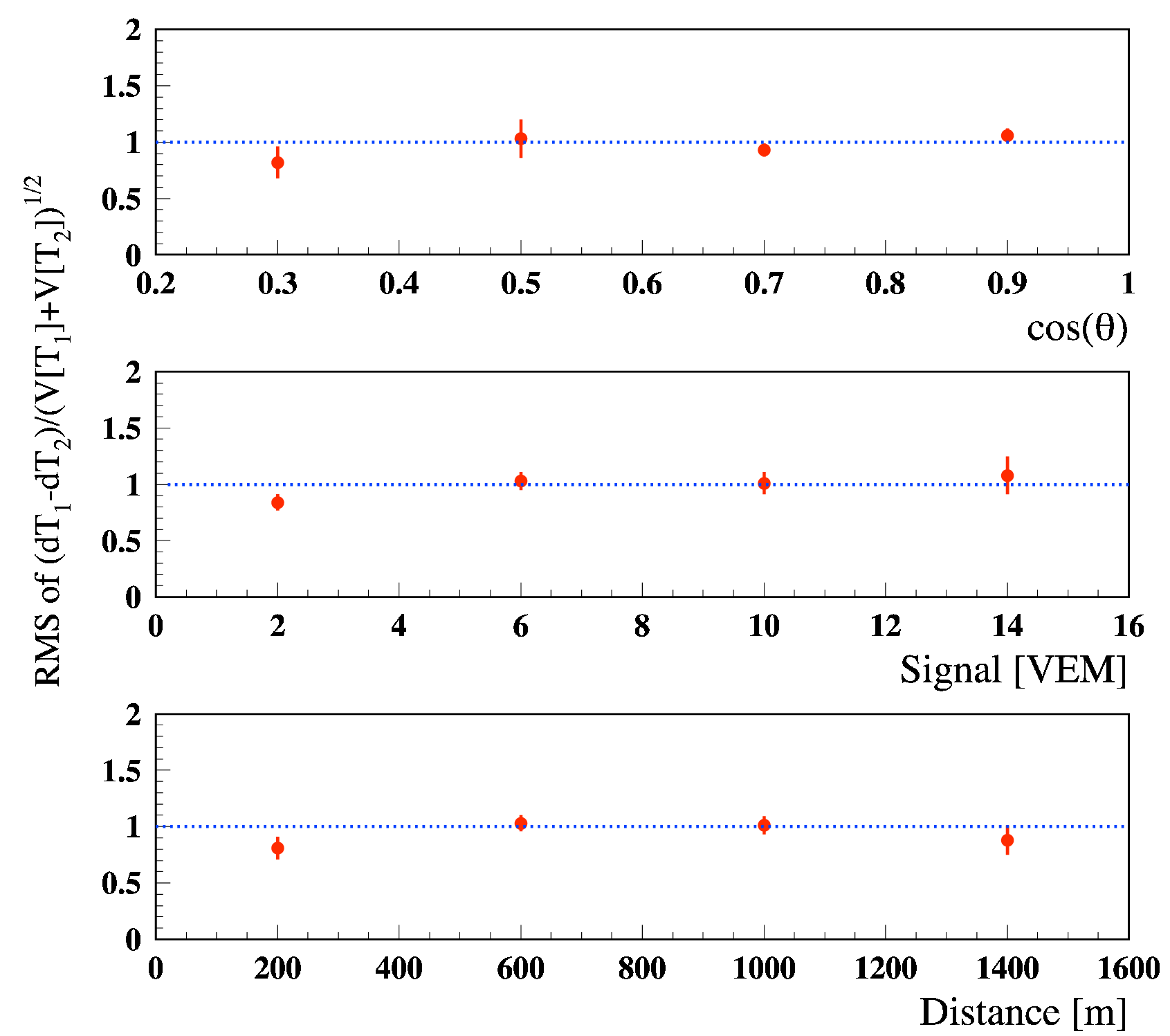}
  \vspace{-0.5cm}
  \caption {The RMS of the distribution of $\Delta T/\sqrt{V[\Delta T]}$,
          as a function of the shower zenith angle (top), the average
          signal in the doublet detectors (middle), and the distance to
          the shower core (bottom). \label{fig:doublets}}
\end{figure}
\begin{figure}[!t]
    \includegraphics*[width=0.45\textwidth]{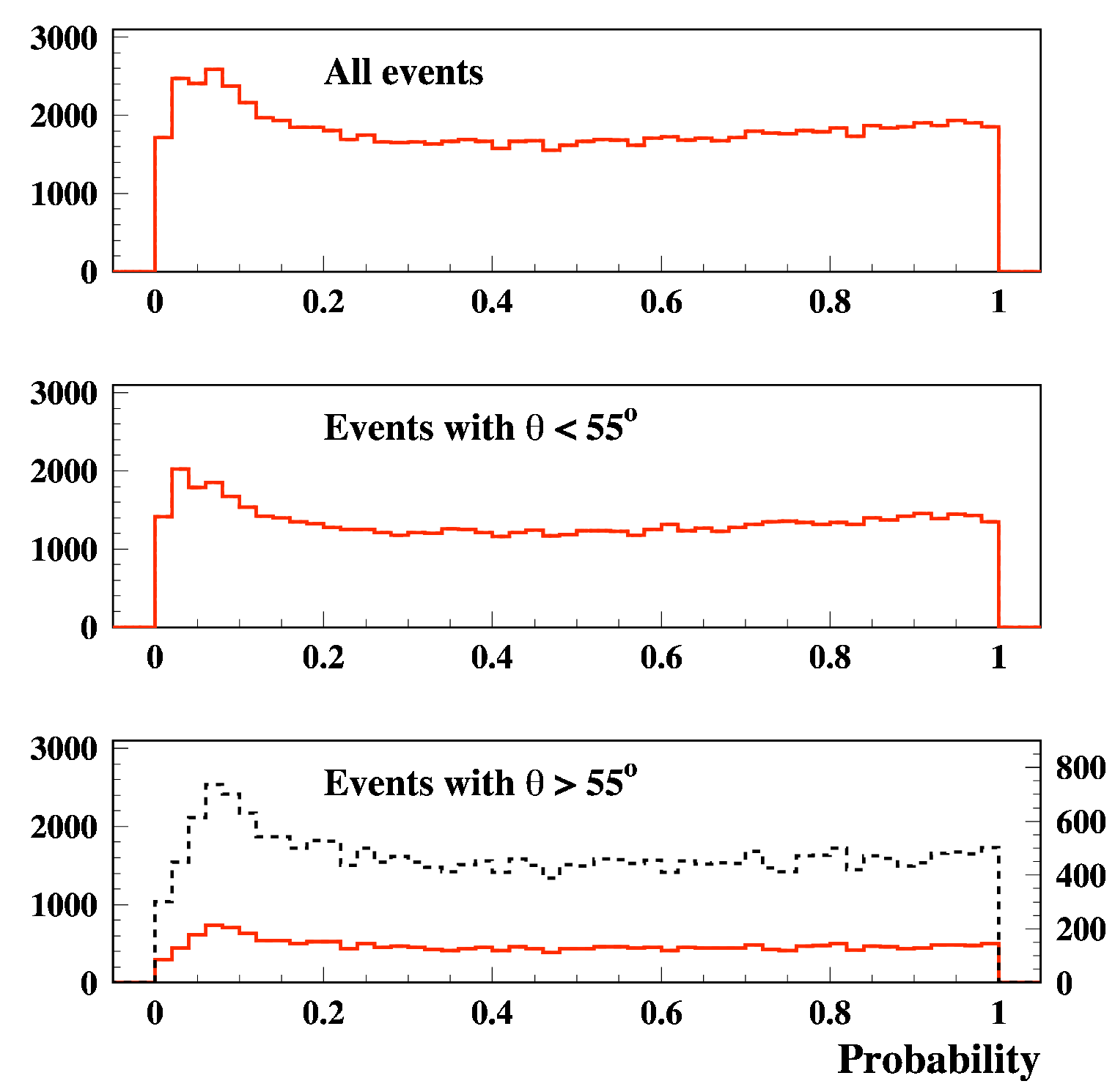}
  \vspace{-0.5cm}
  \caption {The $\chi^2$ probability distribution for all events
    (top), events with zenith angle smaller than 55$^\circ$
    (middle), and events with zenith angle larger than 55$^\circ$
    (bottom). In the last figure the distribution is plotted with two
    different scales, the same than the others (full line)
    for comparison reasons and a zoom (dashed line) to see
    the details. \label{fig:proba}}
\end{figure}

\subsection{Validation of the Time Variance Model}\label{prob}
If the time variance model describes correctly the measurement
uncertainties, the distribution of $\Delta T/\sqrt{V[\Delta T]}$, where
$V[\Delta T]~=~V[T_1^{(1)}]~+~V[T_1^{(2)}]$, should have unit variance.
In figure~\ref{fig:doublets} we show the RMS of the distribution of
$\Delta T/\sqrt{V[\Delta T]}$ for the doublets as a function of
$\cos(\theta)$ (top), the average signal (middle), and the distance
to the core position (bottom). In all the cases, the RMS is almost constant
and close to unity.

In figure~\ref{fig:proba} we plot the $\chi^2$ probability
distribution from our geometrical reconstruction procedure using 
as timing errors the model above. 
We only plot probabilities larger than 1\% to avoid the large peak
at zero corresponding to badly reconstructed events ($\sim$~9\%).
These distribution are almost flat indicating that our error model is correct.

\subsection{Surface Detector Only}\label{SDonly}
Given the correctness of our error model, we can extract
 the angular resolution on an event by event basis directly 
out of the minimization procedure. 

\begin{figure}[t]
    \includegraphics*[width=0.45\textwidth]{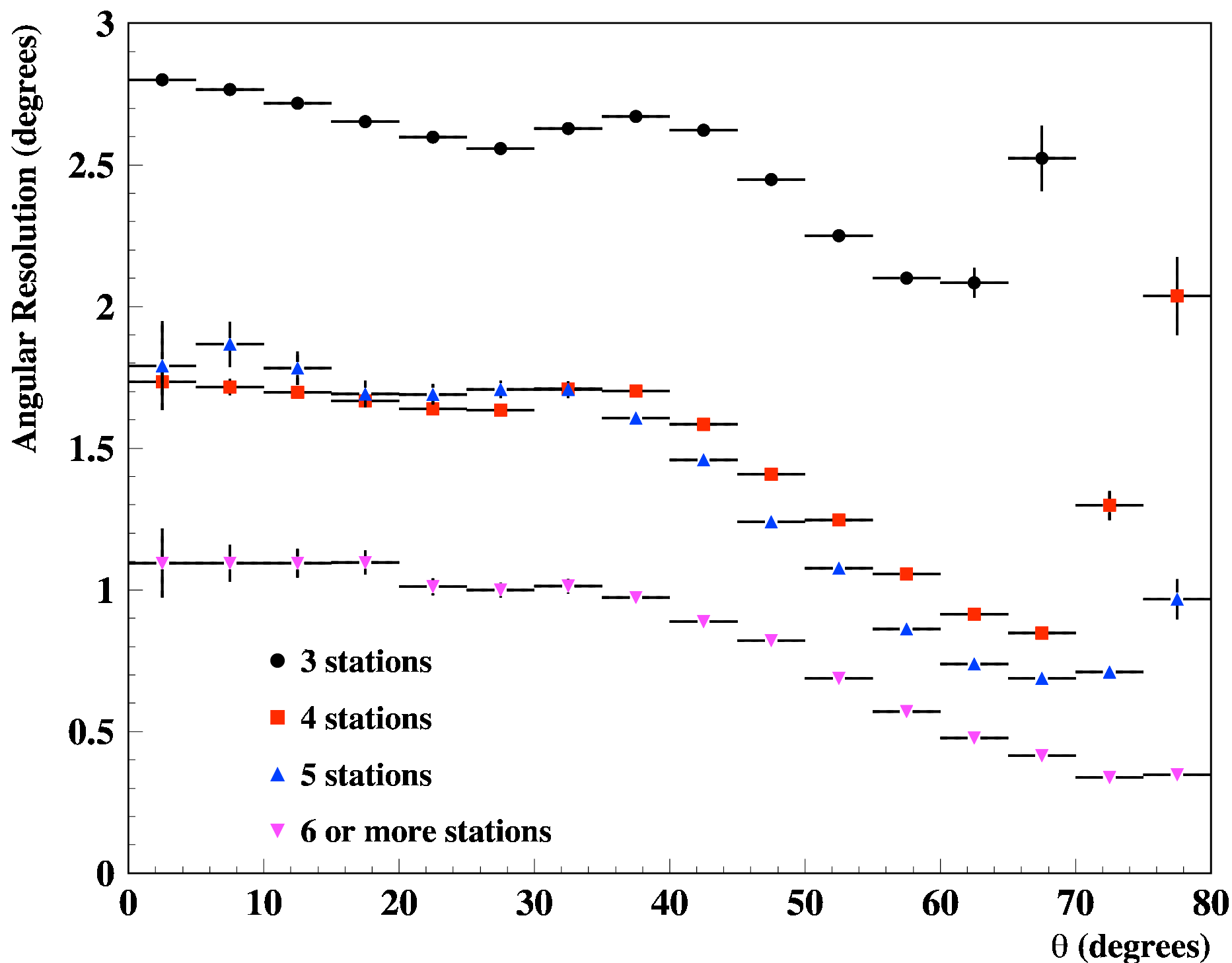}
  \vspace{-0.5cm}
  \caption { Angular resolution ($AR$) for the SD as a function of the
            zenith angle ($\theta$) extracted from l
            reconstruction procedure. The $AR$ is plotted for various
            stations multiplicities (circles: 3 stations, squares: 4
            stations, up triangles: 5 stations, and down triangles: 6
            or more stations).\label{fig:ar}}
\end{figure}

The angular resolution is about $2.7^\circ$ in the worst case of vertical
showers with only 3 stations hit. This value improves significantly
for 4 and 5 stations\footnote{4 and 5 stations events have the same
number of degrees of freedom in the geometrical fit, hence similar resolution.}. 
For 6 or more stations, which corresponds to
events with energies above 10 EeV, the angular resolution is in
all cases better than $1^\circ$. 

All quoted errors are statistical only. We
did not, at this stage, investigate possible biases or systematics
in the determination of the arrival direction angles.

\subsection{Comparison with Hybrid events}
\begin{figure}[t]
    \includegraphics*[width=0.45\textwidth]{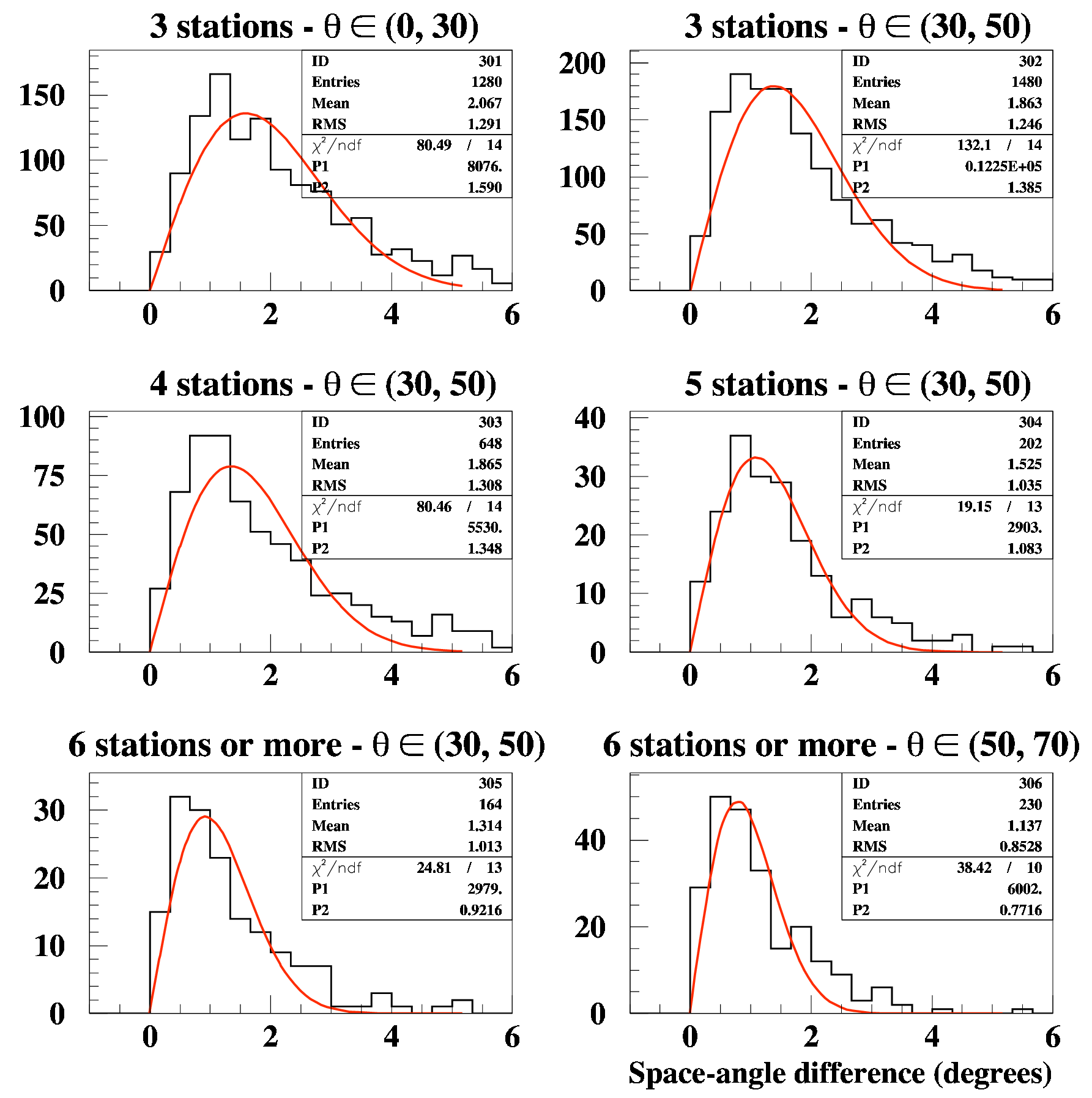}
  \vspace{-0.5cm}
  \caption {Comparison between hybrid and SD-only geometrical
  reconstructions. Top, for 3 stations with two zenith angle ranges
  (0$^\circ<\theta<30^\circ$ and 30$^\circ<\theta<50^\circ$). Middle
  4 stations (left) and 5 stations (right) with
  30$^\circ<\theta<50^\circ$. Bottom, for 6 stations or
  more with two zenith angle ranges (30$^\circ<\theta<50^\circ$ and
  50$^\circ<\theta<50^\circ$).\label{fig:hyb}}
\end{figure}

Finally, in figure \ref{fig:hyb} we show the space angle between the
SD-only and hybrid geometrical reconstructions for showers with
different number of stations and different zenith angle ranges. The
distributions plotted were fitted with a Gaussian resolution function
$dp~\propto~e^{-\eta^2/2\sigma^2}~d($cos$(\eta))d\phi$, where $\eta$
is the space-angle. The $\sigma$ obtained in the fit is related to the
angular resolution by $RA = 1.5~\sigma$.

The parameter $\eta$ given by the fit is in good agreement with the value
from figure\ref{fig:ar} using an hybrid resolution parameter $\eta$ 
of $0.5-0.6^\circ$.

The angular resolution of the surface
detector is then found  to be better than 2.7$^\circ$ for 3-fold events,
better than 1.7$^\circ$ for  4-fold and 5-fold events and better
than 1.0$^\circ$ for higher multiplicity (which corresponds to
energies larger than 10~EeV).

\section{Coverage}
To study anisotropy, one needs the background expectations for the different sky directions under the
assumption of an isotropic CR distribution. 
Modulations of the exposure in right ascension  are induced by 
the dead time of the detectors and by the constantly growing array size. 
So those experimental issues (that are carefully recorded by the auger data 
acquisition system) must be fully taken into account. Also weather variations, 
 may be non-negligible since they affect the shower development in the
atmosphere and/or the response of the  electronics.
 
Preliminary studies of these effects have shown  that the
possible weather-induced background modulations for the present data set
are at a level of 1\%. This is below the Poisson noise for the angular windows
considered\footnote{A detailed account of weather effects is certainly 
necessary to test
  large scale patterns at the few percent level. Relevant studies are in
  progress. }.

We have followed two different technics  \cite{ha05} 
to estimate the coverage for the SD analysis:
\begin{itemize}
\item {\bf The semi-analytic technique:}  The zenith angle distribution 
of the events in the considered energy range is fitted
  and then convoluted with the number
  of  hexagons with active detectors 
(which gives a measure of the aperture
   for events 
   satisfying the quality trigger criterion) as a function of time,
  assuming a uniform response in azimuth. Through this procedure one obtains
  an exposure which accounts
  for the non-saturated acceptance effects and for the non-uniform
  running times and array growth. 

\item {\bf The shuffling technique:} Here the expected number of events in any
  direction is obtained by averaging many data sets obtained by shuffling the
  observed events in the energy range of interest so that the arrival times
  are exchanged among them and the azimuths are drawn uniformly. The
  shuffling can be performed in separate zenith angle bins or  by  just
mixing them
  all, and we found no significant difference between these two
  possibilities. By construction, 
this exposure preserves exactly the $\theta$ distribution of
  the events and accounts for the detector dead times, array growth and even
  for weather-induced modulations. It might however partially
  absorb modulations induced by large scale intrinsic anisotropy present 
in the CR flux, such as those due to a global dipole. 
\end{itemize}

The background estimate obtained with the two technique differ by less than 1\% 
in the GC region. It is  0.5\% larger with shuffling than the  semi-analytic
method. This difference is much smaller than the size of the
excesses we are testing and also below the Poisson
fluctuations, 
In the following we mainly quote the values obtained using the semi-analytic technique.

\section{Data set}
The Auger surface detector \cite{ea04}, 
  has been growing in size during the data taking
period considered in this work.
154 detectors deployed in January 2004 and 
 930 in  March~$30^{th}$ 2006 

The basic cell of surface detectors  is triangular,  with separations of 1.5~km between
detector units, and hence 
the complete array with 1600 detectors will  cover an
area of 3000~km$^2$.

We consider the events from 
the surface detector (SD) array with three or more tanks
triggered in a compact configuration.  The events have
 to satisfy the level 5 quality trigger condition, which requires
 that the detector 
with the highest signal be surrounded by a hexagon of working detectors,
since this ensures that the
 event is well reconstructed. We also restrict the events
to zenith angles $\theta<60^\circ$.

The energies are obtained using the inferred signal size at 1000~m 
 from the
reconstructed shower core, $S(1000)$,
 adopting a conversion that leads to a constant flux
in different sky directions above 3~EeV, where the acceptance is saturated.
This is the so-called Constant Intensity Cut criterion implemented in
\cite{so05}. A calibration of the energies is performed 
using clean fluorescence data.
 The estimated systematic 
uncertainty in the reconstructed shower
 energy with the fluorescence technique is currently 25\% \cite{be05}. 

 From the uncertainty in the measurements of the signals from the
 Cherenkov tanks \cite{piera} 
the statistical uncertainty in the energy  determination which results from the
fitting procedure  is about 20\% for
the energy range considered  in this work, i.e. $10^{17.9}\
{\rm eV}<E<10^{18.5}$~eV.
 Notice that in this energy range
 48\% of the events involve just three tanks, 34\%
involve 4 tanks and only 18\%  more than 4  tanks. For three tank events the 
 68\% quantile  angular resolution is about  
$2.2^\circ$ and the resolution improves for events with 4 tanks or more
\cite{angres}.

Regarding the hybrid events, i.e. those with signal from 
 both the  fluorescence detectors 
(FD) and  surface array,  the 
angular resolution achieved is much smaller, typically below 1 degree
 \cite{angres}.  
Also, given that hybrid events may trigger with just one surface detector, 
the associated energy threshold ($\sim
10^{17}$~eV) is lower, and events up to zenith angles of
 $75^\circ$ are included in the data set.  However, the statistics accumulated
are significantly less, in part due to the $\sim 15$\% duty cycle of the
 fluorescence telescopes and
also because at EeV energies the FD is not fully efficient at
detecting showers over the full SD array.

There are for instance 79265 SD events in the data set considered
with energies  $10^{17.9}\ 
{\rm eV}<E<10^{18.5}$~eV, 
while the corresponding number of well reconstructed hybrid events 
in the same energy range is just 3439.
Note that $\sim 25$\% of the hybrid events in this energy range
involve less than three surface detectors, and are hence not included
in the SD only data set. 

\section{Results}
All results presented in this section are part of a dedicated publication\cite{gc-paper}

\section{AGASA and SUGAR excesses}
\begin{figure*}[ht]
    \includegraphics*[width=0.99\textwidth]{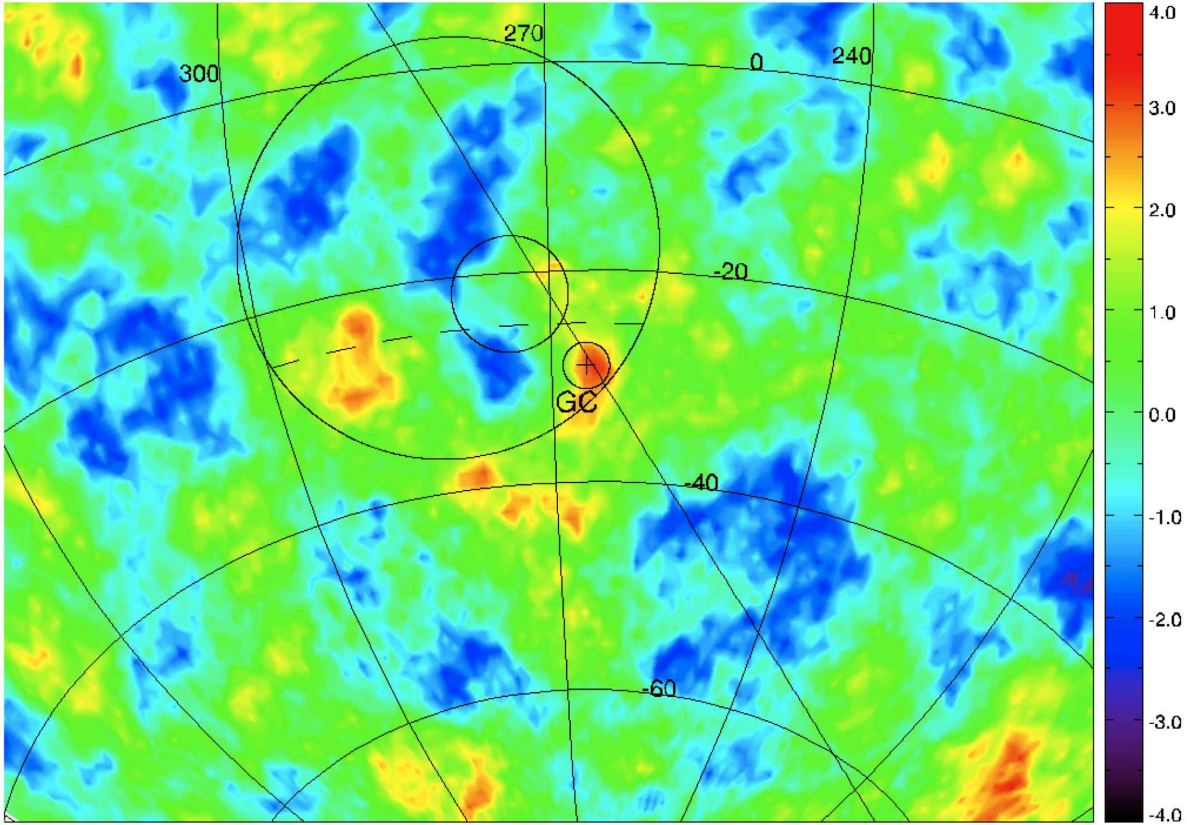}
\caption{Map of CR overdensity significances near the GC region on top-hat
  windows of $5.5^\circ$  radius. The GC location 
is indicated with a cross, the galactic plane as solid line.}
\label{fitdip}
\end{figure*}

The map of overdensities around the GC region is shown on Figure~\ref{fitdip}.
Significances in this map are calculated in circular windows of $5.5^\circ$ using a 
Gaussian approximation to the Poisson law (counts in the $5.5^\circ$ windows are well above 100)
The angular scale chosen and the energy range shown ($10^{17.9}$--$10^{18.5}$~eV) mimics the Sugar search window
and  is also convenient to visualize the distribution of
overdensities in the window explored by  AGASA.
The galactic plane is shown
as a solid line (a cross for the GC)
The big circle region in which AGASA reported an excess (in a slightly narrower
energy range). The
dashed line indicates the lower boundary of their observations.

\begin{figure}[ht]
    \includegraphics*[width=3.in]{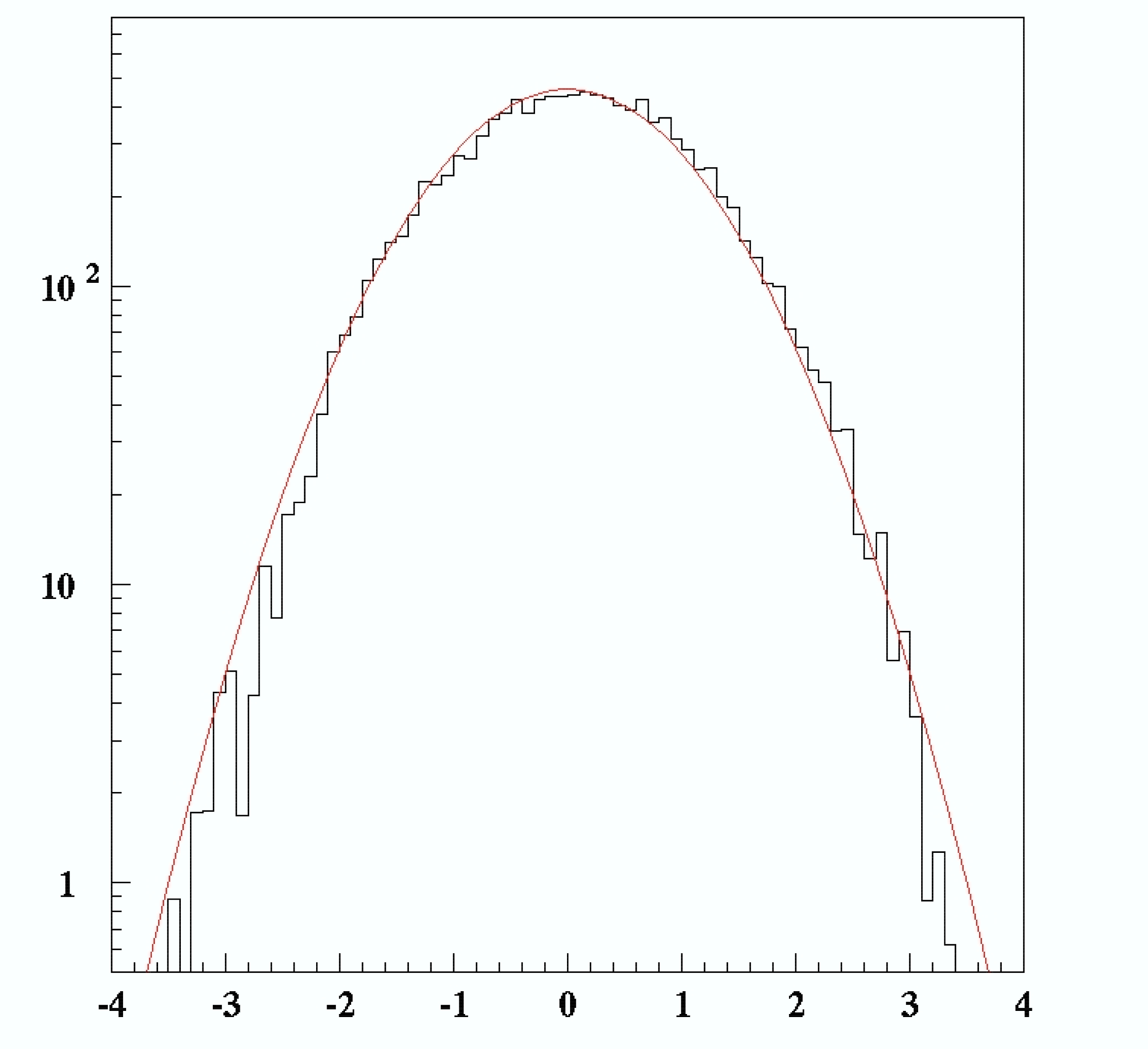}
\caption{Histogram  of overdensities on $5.5^\circ$
radius windows and for  $10^{17.9}\ 
{\rm eV}<E<10^{18.5}$~eV, together with the average isotropic expectations (gaussian of unit width and zero mean). 
The histrogram is build over the same sky patch as figure 6 ($-65^\circ<dec<5^\circ$  and $160^\circ<RA<360^\circ$).}
\label{ovhisto}
\end{figure}

Overdensities present in this map
are consistent with the  expectation from statistical fluctuations of an isotropic sky. 
On Figure~\ref{ovhisto} we show the distribution of these
 overdensities  together with the expectations from an
isotropic flux and no significant departure from isotropy is observed.

In the $20^\circ$ circle centered at the AGASA location and for energies between
$10^{18}\ {\rm eV}<E<10^{18.4}$~eV, we counted 2116 events while
while 2169.7
 are expected using the shuffling technique ( 2159.6 using the semi-analytic) 
No significant excess is observed. 

There may be systematic differences in the
energy calibration of the two experiments.
To test if these differences could possibly mask the AGASA 
excess, we calculated the observed and expected rates for different
energy ranges in offsets of 0.1 decade 
keeping $E_{max}/E_{min}$ fixed. We have also added a
systematic error of 1\% to the expected rates  to
account for possible weather induced modulations.
We found  no significant excesses in the AGASA region for
any of these cases. In particular, at the $2\sigma$ level (95\% CL) the
excess in this region is always less than 6\%, well below the
 22\% excess reported by AGASA.

The acceptance of Auger in this energy range is not yet saturated. 
it is larger for heavy nuclei than for protons because
showers initiated by heavier primaries develop earlier and 
are hence more spread out at ground level. 
Using ref~ \cite{all05} for the acceptance of p and Fe primaries,
we find that the sensitivity to protons is about $\sim 30$\% smaller 
than to Fe in the energy range studied (assuming an $E^{-3}$ spectrum).
If the 22\%  excess reported by AGASA (which had full efficiency at EeV energies) 
was due to nucleons while the background was due to heavy nuclei, at least a 15\%
excess should have been expected in Auger data. This  is 
much larger than the upper limit we are obtaining.

For the excess reported by the  SUGAR experiment, we find in the 
same angular window and energy range that 
$n_{obs}/n_{exp}=286/289.7=0.98\pm 0.06$, 
and hence with 10 times the  statistics we found no no significant excess
in this window. 

\section{Neutron source at the GC}

\subsection{Surface detector}

The energy range for this search is between $E_{min}=10^{17.9}$~eV
and $E_{max}=10^{18.5}$~eV. Below $E_{min}$
 the Auger SD acceptance is  suppressed and most neutrons from 
a source at the GC  would have decayed before reaching the Earth.
Energies above $E_{max}$ are assumed to be  be hard to achieve for
galactic sources.

The optimal  search window for a point-like source is obtained using a Gaussian 
filter matching the angular resolution of the experiment \cite{bi05}.
Following section~\ref{SD-angres}
we used  a 
68\% quantile $\sigma$ of $2.25^\circ$
for the average resolution of the events  in this energy range (in fact more than 80\% of the events have a better 
resolution than this).

In the direction of Sagittarius $A^*$ we get
$n_{obs}/n_{exp}= 53.8/45.8$ (a ratio of $1.17\pm 0.10$).
Applying the results of  \cite{bi05}, we
get a 95\% CL upper bound on the number of events from the source 
of $n_s^{95}=18.5$.

Assuming :
\begin{itemize}
\item that the source spectrum shape is similar to that of the bulk CRs,
($\propto E^{-3}$ in this energy range),
\item that both the bulk CRs and the source CR are protons in this energy range
(We discuss how the limit is modified if the bulk CRs were heavier) ,
\end{itemize}
we have~\cite{bi05} :
\begin{equation}
\label{ratio}
4\pi\eta^2\frac{\Phi_{CR}}{n_{exp}} = \frac{\Phi^{95}_{s}}{n_s^{95}}
\end{equation}
Where $\Phi_{CR}$ and $\Phi^{95}_{s}$ are the bulk CR flux and the source 95\% flux upper limit
integrated in the  energy range under study and where $n_{exp}$ and $n^{95}_s$ are as above.
$\eta$ is the parameter
of the optimal Gaussian window that corresponds to our angular resolution 
\footnote{Remember that $\eta$ and the 68\% resolution quantile $\sigma$
are related by $\sigma=1.5\eta $ (cf section~\ref{SD-angres}).}
(here $\eta = 2.25/1.5 = 1.5^\circ$)

For the differential CR spectrum flux we take
\begin{equation}
\Phi_{CR}(E)\simeq 
\xi\; 30 \left(\frac{E}{\rm
  EeV}\right)^{-3}\!\!\!\!\!\!\!{\rm EeV^{-1}km^{-2}yr^{-1}sr^{-1}},
\end{equation}
which has an $E^{-3}$ dependence. 
The factor $\xi$ is close to unity  and parametrizes the
uncertainties in the CR flux normalization, so that the flux 
bounds will be simply proportional to $\xi$.

From eq.~\ref{ratio} we  get a
95\% CL  upper bound 
for the source flux 
integrated over the
energy range considered of :  
\begin{equation}
\Phi_s^{95}= \xi\; 0.08 \ {\rm km^{-2}\ yr^{-1}}.
\label{sdbound}
\end{equation}

If the bulk of the 
CRs were heavy nuclei the upper limit to the flux from the source
need to to be scaled by a factor $\sim
1.3$ under the assumption that the CRs are iron nuclei and that the source 
is of neutrons.  We thus see
 that the bound on the neutron flux could be up to 
$\sim 30$\% higher if the CR composition at EeV energies were
 heavy.

Theoretical predictions for neutron
fluxes (only those associated with the AGASA claim) are based on
 the AGASA normalization.  This normalization is about a
factor of 3 larger than the Auger one so these predictions 
 must be reduced by  this factor to be compared with our
flux bound. 
 The predictions of refs.~\cite{bo03}, \cite{ah05}
 and  \cite{cr05}, which exceed the
upper-bound obtained by more than one order of magnitude, 
 are already excluded, and that of \cite{gr05} is 
at the level of the present Auger sensitivity.

\subsection{Hybrids}
As discussed earlier, hybrid events,  detected by both
the FD and SD have a better angular resolution, 
$ 0.7^\circ$ at 68\% C.L. in the energy range studied. 

For energy between  $10^{17.9}\ {\rm eV}<E<10^{18.5}$~eV,
 no significant excess is seen in the GC direction. 
Using, for instance an optimal top-hat window of
$1.59\sigma\simeq 0.75^\circ$ radius, 
 0.3 events are expected (shuffling technique estimate)  while 
no event direction falls within that circle.

This gives to a source flux upper-bound at 95\% CL of
\begin{equation}
\Phi_s^{95}=\xi\; 0.15 \ {\rm km^{-2}\ yr^{-1}},
\end{equation}
which is about a factor of two weaker than the SD flux bound.

Note that quality 
requirement for hybrid events is to have the maximum of the shower
development inside the field of view of the telescopes, this affects
the sensitivity to different primaries.
The bound obtained here is a conservative one if the bulk of the CRs are
heavy nuclei.

\section{Conclusions}
The Pierre Auger Observatory is a hybrid detector with excellent capabilities
for studying the highest energy cosmic rays.
Much of its capability stems from the accurate geometric reconstruction it achieves.

The construction of the Southern Observatory is well under way.
Eighteen FD telescopes and more than $60\%$ of the surface array are in operation
taking data routinely.
At the present rate of deployment,
construction will be finish in mid 2007.
Detectors are performing very well and the first results are very encouraging.

The angular resolution of the surface
detector was found to be better than 2.7$^\circ$ for 3-fold events,
better than 1.7$^\circ$ for  4-fold and 5-fold events and better
than 1.0$^\circ$ for higher multiplicity (which corresponds to
energies larger than 10~EeV). 

Using a bit more than the 2 first year of  Auger data we have searched for localized
anisotropy near
the direction of the Galactic Centre.
With statistics much greater than those of
previous experiments, we have looked for a point-like source in the
direction of Sagittarius $A$, without finding a significant excess.

We exclude several scenarios of neutron sources in the GC as suggested recently.
 Our searches  do not support the large excesses reported in AGASA
data (of 22\% on $20^\circ$ scales) and SUGAR data 
(of 85\% on $5.5^\circ$ scales).

\section{Acknowledgments}

This work was supported by the Conselho Nacional de
Desenvolvimento Cient\'{\i}fico e Tecnol\'ogico (CNPq), Brazil,  and the
Centre National de la Recherche Scientifique, Institut National de
Physique Nucl\'eare et Physique des Particules (IN2P3/CNRS), France.

\end{document}